\pdfoutput=1

\documentclass[a4paper,12pt]{iopart}

\usepackage[]{amssymb}
\usepackage[]{amsthm}
\usepackage[]{mathrsfs}
\usepackage[]{tensor}
\usepackage[]{bm}
\usepackage[]{eucal}
\usepackage[unicode]{hyperref}
\usepackage{xcolor}
\definecolor{pastblue}{HTML}{336699}
\definecolor{linkcol}{HTML}{663333}
\hypersetup{colorlinks,linkcolor={linkcol},citecolor={pastblue},urlcolor={pastblue}}

\newcommand{\dd}{\partial}
\newcommand{\df}{\mathrm{d}}
\newcommand{\w}{\wedge}
\newcommand{\Lie}{\pounds}
\newcommand{\nab}[1]{\nabla_{\!#1}}
\newcommand{\qqd}{\ , \quad}
\newcommand{\bc}{\begin{center}}
\newcommand{\ec}{\end{center}}
\newcommand{\be}{\begin{equation}}
\newcommand{\ee}{\end{equation}}

\theoremstyle{plain} \newtheorem{tm}{Theorem}[]
\theoremstyle{plain} \newtheorem{lm}[tm]{Lemma}
\theoremstyle{definition} \newtheorem{defn}[tm]{Definition}
\newcommand{\btm}{\begin{tm}}
\newcommand{\etm}{\end{tm}}
\newcommand{\blm}{\begin{lm}}
\newcommand{\elm}{\end{lm}}
\newcommand{\bdefn}{\begin{defn}}
\newcommand{\edefn}{\end{defn}}

\begin{document}

\begin{flushright}
ZTF-EP-15-04
\end{flushright}

\note[Does three dimensional electromagnetic field inherit the spacetime symmetries?]{Does three dimensional electromagnetic field inherit the spacetime symmetries?}

\author{M Cvitan$^a$, P Dominis Prester$^b$ and I Smoli\'c$^a$}

\medskip

\address{$^a$ Theoretical Physics Division of Particles and Fields, Department of Physics, Faculty of Science, University of Zagreb, Bijeni\v cka cesta 32, 10000 Zagreb, Croatia}

\smallskip

\address{$^b$ Department of Physics, University of Rijeka, Radmile Matej\v ci\'c 2, 51000 Rijeka, Croatia}

\eads{\mailto{mcvitan@phy.hr}, \mailto{pprester@phy.uniri.hr}, \mailto{ismolic@phy.hr}}

\medskip

\begin{abstract}
We prove that the electromagnetic field in a $(1+2)$-dimensional spacetime necessarily inherits the symmetries of the spacetime metric in a large class of generalized Einstein-Maxwell theories. The Lagrangians of the studied theories have general diff-covariant gravitational part and include both the gravitational and the gauge Chern-Simons terms.
\end{abstract}


\bigskip


\vspace{20pt}

\section{Introduction}

Three-dimensional spacetime is an important playground for various aspects of gra\-vi\-ta\-ti\-onal physics. An example of well-studied gravitational theory in 3d is topologically massive gravity, introduced by Deser, Jackiw and Templeton \cite{DJT82a,DJT82b} which contains higher derivative gravitational Chern-Simons, as well as the gauge Chern-Simons corrections to the Einstein-Maxwell gravity.

\medskip

Here we study a broader class of theories described by the action,
\be
I = \frac{1}{16\pi} \int \left( \mathbf{L}_{\mathrm{g}} + \mathbf{L}_{\mathrm{em}} \right) \ ,
\ee
with the gravitational Lagrangian $\mathbf{L}_{\mathrm{g}}$ and the (minimally coupled) electromagnetic Lagrangian $\mathbf{L}_{\mathrm{em}}$. The corresponding gravitational field equations have the form
\be\label{eq:ET}
E_{ab} = 8\pi T_{ab} \ ,
\ee
where the Lagrangian $\mathbf{L}_{\mathrm{g}}$ is such that the tensor $E_{ab}$ is allowed to be diff-covariant and differentiable but otherwise completely general function of the spacetime metric, the Ricci tensor\footnote{In three-dimensions the Riemann tensor is not independent but can be written in terms of the Ricci tensor and the metric tensor.}, the Levi-Civita tensor and covariant derivatives. The electromagnetic Lagrangian $\mathbf{L}_{\mathrm{em}}$ consists of the conventional Maxwell term and the gauge Chern-Simons term (with the coupling constant $\mu$),
\be\label{gaugeL}
\mathbf{L}_{\mathrm{em}} = -\frac{1}{2} \left( F \w {*F} + \mu F \w A \right) \ .
\ee
It is important to emphasize that the Maxwell Chern-Simons term doesn't depend on the spacetime metric, hence it doesn't change the form of the electromagnetic energy-momentum tensor,
\be\label{Tem}
T_{ab} = \frac{1}{4\pi} \left( F_{ac} \tensor{F}{_b^c} - \frac{1}{4}\,g_{ab} F_{cd} F^{cd} \right) \ .
\ee
The Maxwell-Chern-Simons field equations are given by
\be\label{eq:MaxwCS}
\df F = 0 \qqd \df\,{*F} = -\mu F \ .
\ee

For example, in the topologically massive gravity the tensor $E_{ab}$ is the sum of the Einstein tensor $G_{ab}$, the cosmological constant term and the Cotton tensor,
\be\label{TMG}
E_{ab}^{\mathrm{TMG}} = G_{ab} + \Lambda g_{ab} + \lambda C_{ab} \ ,
\ee
where $\lambda$ is the gravitational Chern-Simons coupling constant and
\be\label{Cotton}
C_{ab} = \tensor{\epsilon}{_a^c^d} \nab{c} \left( R_{db} - \frac{1}{4}\,R g_{db} \right) \ .
\ee
Other additional terms have been introduced and analysed within the ``new massive gravity'' model \cite{BHT09} and its extensions \cite{Sinha10}. Some exact solutions of the field equations (\ref{eq:ET}) and (\ref{eq:MaxwCS}) with $E_{ab}$ and $T_{ab}$ given by (\ref{TMG}) and (\ref{Tem}) were found and analysed in \cite{DS00,MCGL08}. An example of non-polynomial tensor $E_{ab}$ appears in the Born-Infeld gravity model \cite{DG98,BF10,GST10}.

\bigskip

Our focus is on spacetimes $(M,g_{ab},F_{ab})$ which admit at least one (sufficiently smooth) Killing vector field $\xi^a$, such that $\Lie_\xi g_{ab} = 0$. A typical (and often implicit) assumption in the literature is that the gauge fields \emph{inherit} the symmetries, $\Lie_\xi F_{ab} = 0$. 
For example in \cite{BCPDPS11b}, using this as an assumption it was shown that in spherically symmetric spacetimes general, $D \geq 3$, Chern-Simons terms \cite{BCPDPS11a,BCPDPS13} do not influence equations of motion. However, there are known examples of 4-dimensional spacetimes, solutions to the Einstein-Maxwell equations, where such assumptions do not hold \cite{MW75,SKMHH}. Several analyses \cite{Woo73a,Woo73b,MzHRS74,MW75,WY76b,Tod06}, focused on four dimensional electromagnetic fields, have shown that in general the symmetry inheritance is broken in the following way
\be
\Lie_\xi F_{ab} = \beta\,{*F}_{ab}
\ee 
with some function $\beta$, which is a constant if $F_{ab}$ is non-null (we say that $F_{ab}$ is a null electromagnetic field if $F_{ab}F^{ab} = F_{ab}{*F}^{ab} = 0$). As $F_{ab}$ and its Hodge dual ${*F}_{ab}$ are both 2-forms only in 4-dimensional spacetimes, it is not straightforward to extrapolate this conclusion to higher or lower dimensional cases. Our aim is to fill this gap in the literature, at least for the three-dimensional case.

\bigskip

\section{Symmetry inheritance}

Clearly, for any Killing vector field $\xi^a$ we immediately have $\Lie_\xi E_{ab} = 0$. By using the field equation (\ref{eq:ET}) it follows that
\be\label{eq:LieT}
\Lie_\xi T_{ab} = 0 \ .
\ee
The main idea is to use the above equation to conclude as much as possible about the symmetries of the electromagnetic field, described by the tensor field $\Lie_\xi F_{ab}$. 

\medskip

To begin with, it is convenient to split $F_{ab}$ into electric and magnetic parts.
Given a non-null vector field $X^a$ we introduce formal electric 1-form $E_a$ and magnetic scalar $B$ \cite{Heusler,ISm12,ISm14},
\be\label{eq:EB}
E \equiv -i_X F \qqd B \equiv i_X {*F} 
\ee
which allow us to make decomposition of the electromagnetic 2-form,
\be\label{eq:FEB}
-N F = X \w E + *(BX)
\ee
where $N \equiv X^a X_a$ is the norm of the vector field $X^a$. In some cases these fields can be directly related to physical observables: An observer with 3-velocity $u^a$ will measure the electric and magnetic field given by the choice $X^a = u^a$. On the other hand, in the context of symmetry analysis, a natural and practical choice is a decomposition (\ref{eq:FEB}) with respect to the Killing vector field, $X^a = \xi^a$. 

\medskip

We now use this symmetry motivated decomposition on the right hand side of the expression for the energy momentum tensor (\ref{Tem}). We have
\be\label{eq:Txixi}
8\pi T_{ab}\xi^a\xi^b = E_a E^a + B^2 
\ee
\be\label{eq:xiTxi}
4\pi *(\xi \w T(\xi))_a = -B E_a
\ee
where $T(\xi)_a \equiv T_{ab}\xi^b$. The Lie derivatives $\Lie_\xi$ of the left hand sides of both of these equations vanish due to (\ref{eq:LieT}), implying in turn that
\be\label{eq:LieTxixi}
E^a \Lie_\xi E_a + B \Lie_\xi B = 0
\ee
and
\be\label{eq:LiexiTxi}
B \Lie_\xi E_a + (\Lie_\xi B) E_a = 0 \ .
\ee
There is, however, a potential technical issue if the Killing vector field $\xi^a$ becomes null on some subset of the spacetime: as it is no longer to possible to ``reconstruct'' $F_{ab}$ from $E_a$ and $B$ at such points, one has to find another appropriate vector field. Let us denote by $Z \subseteq M$ the (closed) set where $\xi^a \xi_a = 0$. We always assume that all points where the Killing vector field vanishes, $\xi^a = 0$ (e.g.~the axis of symmetry or the bifurcation surface of the Killing horizon) belong to the boundary $\dd Z$. Since the norm of the Killing vector is constant along its orbits (integral curves), it follows that each orbit of $\xi^a$ is either contained in $Z$ or disjoint from it. On each orbit $\gamma$ of $\xi^a$ contained in the interior $Z^\circ$, at a point $p \in \gamma \subseteq Z^\circ$ one can choose an auxiliary timelike vector $v^a$ and then Lie drag it with respect to $\xi^a$ along the $\gamma$. Furthermore, let $\Sigma_p$ be a (locally defined) smooth spacelike hypersurface passing through the point $p$ and $\mathscr{U}_p \subseteq Z^\circ$ a neighbourhood of the point $p$. Then, starting with a choice of (sufficiently smooth) timelike vector field $v^a$ on the intersection $\Sigma_p \cap \mathscr{U}_p$, one can repeat the procedure of Lie dragging along each orbit of $\xi^a$ which intersects the set $\Sigma_p \cap \mathscr{U}_p$. This allows us to construct a timelike vector field $v^a$ which immediately satisfies $\Lie_\xi v^a = 0$ on a tubular neighbourood of the orbit $\gamma$ and can be used there for the decomposition (\ref{eq:FEB}). In order to avoid confusion, let us introduce the notation
\be
\widetilde{E} = -i_v F \qqd \widetilde{B} = i_v {*F} 
\ee
for the electric and the magnetic fields defined with respect to the vector field $v^a$. By construction we have that $\Lie_\xi (T_{ab} v^a v^b) = 0$ and $\Lie_\xi *(v \w T(v)) = 0$, so that equations analogous to (\ref{eq:LieTxixi}) and (\ref{eq:LiexiTxi}) follow from here,
\be\label{eq:LieTvv}
\widetilde{E}^a \Lie_\xi \widetilde{E}_a + \widetilde{B} \Lie_\xi \widetilde{B} = 0 \ ,
\ee
\be\label{eq:LievTv}
\widetilde{B} \Lie_\xi \widetilde{E}_a + (\Lie_\xi \widetilde{B}) \widetilde{E}_a = 0 \ .
\ee
We can now state and prove the main result of this paper.

\medskip


\btm
Let $(M,g_{ab},F_{ab})$ be a $(1+2)$-dimensional spacetime, solution to (\ref{eq:ET}) and (\ref{eq:MaxwCS}) with (sufficiently smooth) Lorentzian metric $g_{ab}$ and electromagnetic field $F_{ab}$, and allowing a (sufficiently smooth) Killing vector field $\xi^a$. Then the symmetry is necessarily inherited,
\be
\Lie_\xi F_{ab} = 0 \ .
\ee
\etm


\medskip

\noindent
The gist of the proof is to split the spacetime into four regions corresponding to points where $B$ and $\xi^a \xi_a$ are zero/non-zero, then to do an analysis on interiors of these regions, and finally to extend the conclusions to the boundary points of each region. In order to extend to the boundary we need to assume the continuity of the fields and use the following two elementary results from the point-set topology.

\medskip

\blm
Let $X$ be a topological space, $Y$ a Hausdorff topological spaces, and $f,g : X \to Y$ continuous maps. If $S \subseteq X$ is a set such that $f(a) = g(a)$ for all $a \in S$, then this is also true at all points of its closure $\overline{S}$.
\elm

\medskip

\noindent
The second lemma ascertains that all the points of the manifold are covered by the end of the process.

\medskip

\blm
Let $X$ be a topological space and $A \subseteq X$ its open or closed subset. Then the boundary $\dd A$ is a closed, nowhere dense set and the closure of its complement is the whole space $X$,
$$\overline{X - \dd A} = X - (\dd A)^\circ = X \ .$$
\elm 

\noindent
The set $X$ can represent the whole spacetime manifold $M$ or just some open subset of $M$ with the induced subspace topology.

\medskip

The introduction of the four regions is necessary in order to cover the cases in which, say, region $\xi^a \xi_a = 0$ has a nonempty interior and is not just a lower dimensional subset of $M$. In case the whole region $\xi^a \xi_a = 0$ is just a boundary of an open region $\xi^a \xi_a \neq 0$, the conclusion would follow simply from continuity by extending the results from the open region $\xi^a \xi_a \neq 0$ to the boundary.

\bigskip

\emph{Proof of the Theorem 1}. If we denote the (possibly empty) closed set of points where $F_{ab} = 0$ by $W \subseteq M$, then the claim is trivial on its interior $W^\circ$. Therefore, we focus our discussion on the open set $M - W$ where the electromagnetic field is nonvanishing and extend the conclusions to the boundary $\dd W$ using lemma 2 and the continuity of the tensor field $\Lie_\xi F_{ab}$.

\medskip

The proof rests upon the decomposition (\ref{eq:FEB}) of the electromagnetic tensor $F_{ab}$ with respect to the Killing vector field $\xi^a$. Accordingly, we treat two separate cases:

\bigskip

\begin{itemize}

\item[(1)] The points where $F_{ab} \neq 0$ and $\xi^a \xi_a \neq 0$, i.e.~the points from the open set 
$$O = (M - W) \cap (M - Z) = M - (W \cup Z) \ .$$ 
We consider two separate subcases:

\bigskip

\begin{itemize}

\item[(a)] On the open subset of points where $B \ne 0$ holds, the equation (\ref{eq:LiexiTxi}) implies
\be
\Lie_\xi E_a = -\frac{\Lie_\xi B}{B}\,E_a \ ,
\ee
which together with (\ref{eq:LieTxixi}) gives
\be
\left( B^2 - E_a E^a \right) \Lie_\xi B = 0 \ .
\ee
Thus, on the open subset where $B^2 \ne E_a E^a$ we immediately have $\Lie_\xi B = 0$. On the other hand, within the interior of the closed set where the equality $B^2 = E_a E^a$ holds, the Lie derivative $\Lie_\xi$ of this equality together with (\ref{eq:LieTxixi}) implies again that $\Lie_\xi B = 0$. Furthermore, $\Lie_\xi B = 0$ implies via (\ref{eq:LiexiTxi}) that $\Lie_\xi E_a = 0$. Hence, using the continuity of the field $\Lie_\xi F_{ab}$ we can conclude that the symmetry is inherited, $\Lie_\xi F_{ab} = 0$, on all points of the set $O$ where $B \ne 0$.

\bigskip

\item[(b)] On the interior of the closed set of points where $B = 0$ holds we have
\be
\Lie_\xi {*F} = ( i_\xi \df + \df i_\xi )\,{*F} = -\mu\,i_\xi F
\ee
By taking the Hodge dual (which commutes with $\Lie_\xi$ since $\xi^a$ is a Killing vector field) we get
\be\label{eq:LieFmuF}
\Lie_\xi F = \mu *i_\xi F
\ee
Here it might seem that the presence of the gauge CS term might allow the breaking of the symmetry inheritance, however, we shall show that this cannot happen. Using Maxwell-Chern-Simons equations (\ref{eq:MaxwCS}) and (\ref{eq:LieFmuF}) we have
\be
\df E = -\df i_\xi F = -\Lie_\xi F + i_\xi \df F = -\mu * i_\xi F = \mu\,{*E}
\ee
and then
\be\label{eq:LieEmuExi}
\Lie_\xi E = (i_\xi \df + \df i_\xi)\,E = i_\xi \df E = \mu *(E \w \xi) \ .
\ee 
Let us now look back at the complete electromagnetic energy-momentum tensor (\ref{Tem}), expressed with electric field 1-form (note that by assumption $B = 0$),
\be
4\pi T_{ab} = \frac{1}{N}\,E_a E_b + \frac{E_c E^c}{N^2}\,\xi_a \xi_b - \frac{E_c E^c}{2N}\,g_{ab} \ ,
\ee
where $N = \xi^a \xi_a$. Using the fact that $\Lie_\xi \xi^a = 0$, $\Lie_\xi N = 0$ and $\Lie_\xi (E^c E_c) = 0$ (which follows from (\ref{eq:LieTxixi})), we have
\be\label{eq:LieEE}
0 = 4\pi \Lie_\xi T_{ab} = \frac{1}{N}\,\Lie_\xi (E_a E_b) \ .
\ee
Contracting with $E^b$ we get
\be
E^b E_b \Lie_\xi E_a = 0
\ee
So, at each such point either $\Lie_\xi E_a = 0$, therefore $\Lie_\xi F_{ab} = 0$ and the symmetry is inherited, or $E^a$ is a null vector. Let us look more closely at a point $p\in O$ where the latter case occurs. Here it is easy to see that 
\be
(*(E \w \xi) | *(E \w \xi)) = -(E \w \xi | E \w \xi) = -N E^a E_a = 0 \ ,
\ee
whence $\Lie_\xi E = \mu *(E \w \xi)$ is also null and, furthermore, by (\ref{eq:LieTxixi}), orthogonal to $E^a$. This implies that these two are proportional at the point $p$,
\be
\Lie_\xi E_a = \alpha E_a \ ,
\ee
which together with (\ref{eq:LieEE}) gives
\be
2\alpha E_a E_b = 0 \ .
\ee
Since by assumption $p \notin W$ it follows from here that $\alpha = 0$ and thus $\Lie_\xi E_a = 0$. Note, however, that symmetry inheritance $\Lie_\xi F_{ab} = 0$ contradicts the equation (\ref{eq:LieFmuF}) under the assumption $F_{ab} \ne 0$, unless $\mu = 0$. Again, boundary points of the set where $B = 0$ are covered by the argument of continuity.

\end{itemize}

\bigskip

\item[(2)] The points where $F_{ab} \neq 0$ and $\xi^a \xi_a = 0$, i.e.~the points from the open set $Z^\circ \cap (M - W)$. Here, on a tubular neighbourhood of each orbit of $\xi^a$, one can construct an auxiliary timelike vector field $v^a$ such that $\Lie_\xi v^a = 0$, as described above the theorem 1, and use it for the decomposition (\ref{eq:FEB}). As the equations (\ref{eq:LieTvv}) and (\ref{eq:LievTv}) are completely analogous to the equations (\ref{eq:LieTxixi}) and (\ref{eq:LiexiTxi}), the proof in the (a) case from above (that is, when $\widetilde{B} \ne 0$) can be repeated essentially unaltered. In the (b) case (that is, when $\widetilde{B} = 0$) we turn to the complete energy-momentum tensor
\be
4\pi T_{ab} = \frac{1}{V}\,\widetilde{E}_a \widetilde{E}_b + \frac{\widetilde{E}_c \widetilde{E}^c}{V^2}\,v_a v_b - \frac{\widetilde{E}_c \widetilde{E}^c}{2V}\,g_{ab} \ .
\ee
where $V = v_a v^a$. Using the fact that, by construction, $\Lie_\xi V = 0$ and that (\ref{eq:LieTvv}) implies $\widetilde{E}^c \Lie_\xi \widetilde{E}_c = 0$, we have
\be
\Lie_\xi (\widetilde{E}_a \widetilde{E}_b) = 0 \ . 
\ee
Additional contraction with $\widetilde{E}^b$ gives us
\be
\widetilde{E}^b \widetilde{E}_b \Lie_\xi \widetilde{E}_a = 0
\ee 
So, either $\Lie_\xi \widetilde{E}_a = 0$ and we are finished with the proof, or $\widetilde{E}^b \widetilde{E}_b = 0$. The latter case, however, is impossible since $v^a \widetilde{E}_a = 0$ and $v^a$ is by construction timelike!

\medskip

Finally, the conclusions about the symmetry inheritance can be extended to the set $\dd Z \cap (M - W)$ using continuity of the tensor field $\Lie_\xi F_{ab}$.

\end{itemize}

\qed

\vspace{20pt}

Now, as a consequence of the symmetry inheritance we know that $\df E = 0$ (the electric field 1-form is a closed form) and $\df B = -\mu E$. Note that in the absence of the gauge CS term, $\mu = 0$, the magnetic field $B$ is necessarily constant\footnote{In the case when $\xi^a \xi_a = 0$, $E_a$ and $B$ defined as in (\ref{eq:EB}) do not carry enough information for (\ref{eq:FEB}) to be useful, nevertheless $B$ defined in this way would be constant.}! This, however doesn't imply that the general observer with the 3-velocity $u^a$ will measure a constant magnetic field $\widehat{B}$. We have (see \cite{KS13})
\be
\widehat{B} = i_u {*F} = -\frac{1}{N}\,i_u \Big(\!*(\xi \w E) - B\,\xi\Big) = \frac{1}{N}\,\Big( B\,i_u\xi + *(\xi \w u \w E) \Big)
\ee
For example, for a stationary observer in a stationary spacetime with the corresponding Killing vector $k^a$, its 3-velocity is given by
\be
u^a = \frac{k^a}{\sqrt{-k_b k^b}}
\ee
at all points where $k^a$ is timelike, and the value of the magnetic scalar measured by this observer is 
\be
\widehat{B} = \frac{B}{\sqrt{-k_b k^b}} \ ,
\ee
which is reminiscent of the gravitational redshift and Tolman's law. In this example, the magnetic field $\widehat{B}$ is time independent, but can have different values at different points of the spacelike hypersurfaces.

\bigskip

It is often very practical to introduce the gauge 1-form $A_a$ via $F = \df A$, unique up to a gauge transformation $A' = A + \df\lambda$. What does the Theorem 1 tell us about the symmetry inheritance of $A_a$? As the Lie and the exterior derivative commute, we know that $\Lie_\xi A$ must be a closed form. Thus, the Poincar\'e lemma implies that at least locally there exist a function $\alpha$, such that $\Lie_\xi A = \df\alpha$. But then, using appropriate local choice of the gauge defined by $\Lie_\xi \lambda = -\alpha$, we have $\Lie_\xi A' = \Lie_\xi A + \df \Lie_\xi \lambda = 0$.

\bigskip

A word has to be said on the scope of our theorem. It applies to all theories in three dimensions which satisfy the following conditions: (i) the equation for gravity can be put in the form (\ref{eq:ET}), where the energy-momentum tensor $T_{ab}$ is as in the standard Maxwell theory and $\Lie_\xi E_{ab} = 0$ holds; (ii) equations of motion for the electromagnetic field are as in (\ref{eq:MaxwCS}). We have mentioned that this includes all diff-covariant metric theories of gravity minimally coupled to the electromagnetic field with the Maxwell plus gauge Chern-Simons Lagrangian. Naturally, one may contemplate extending the theorem to other classes of theories. However, under the current proof, this is possible only if the above mentioned requirements are satisfied. For example, if one considers theories which contain additional matter or torsion, one has to assume that additional degrees of freedom do not couple to the electromagnetic field and do not violate symmetry of the tensor $E_{ab}$, which obviously puts some symmetry requirements on these fields. An interesting problem would be to consider the massive Proca theory instead of the massless spin-1 $U(1)$ gauge field. However, addition of mass affects both the equations of motion and the energy-momentum tensor in a way which makes it unclear how to extend the analysis and conclusions presented in this paper.

\bigskip

\section{Final remarks}

We have presented a proof that the question from the title of the paper has an \emph{affirmative} answer for a broad range of three-dimensional theories. For example, the argument works in the case of a ``typical'' spacetime with symmetries, consisting of several regions with non-null Killing vector, parcelled with Killing horizons and pierced with axis of symmetry. The proof may possibly break at points where the Killing vector field $\xi^a$ or the electromagnetic field $F_{ab}$ have discontinuities.

\medskip

The result presented in this paper can be seen as a part of a wider survey of symmetry inheritance for various physical fields. A series of papers \cite{Hoen78,BST10a,BST10b,GJ14b,ISm15} has shown that possible symmetry inheritance breaking is highly restricted for the real and complex scalar fields. For example, in addition to well-known solution by Wyman \cite{Wyman81}, an important example of a rotating black hole with complex scalar field, which evades the Bekenstein's no-hair theorem due to symmetry noninheritance, has been recently found by Herdeiro and Radu \cite{HR14}.

\medskip

Now that we know symmetry inheritance properties of the electromagnetic field in three and four dimensional spacetimes, it remains to be seen what the general conclusions in higher dimensional spacetimes are. Also, apart from one analysis of the electromagnetic field with perfect fluid \cite{WY76a}, there are no other results about the symmetry inheritance when multiple fields are present or when the gauge field is nonminimally coupled to gravity. Another line of pursuit both in three and higher number of dimensions would be to extend the result to gravity theories written in the vielbein formalism.

\bigskip

\section{Acknowledgments}

\medskip

The research has been supported by Croatian Science Foundation under the project No.~8946 and by University of Rijeka under the research support No. 13.12.1.4.05.

\vspace{30pt}

\bibliographystyle{iopart-num}
\bibliography{3DMaxCS}

\end{document}